\documentclass[aps,pre,twocolumn,superscriptaddress,showpacs]{revtex4-1}
\usepackage[dvips]{graphicx}
\usepackage{amsmath}
\begin{document}
\title{Distinct dynamical behavior in Erd\H{o}s--R\'enyi networks, regular random networks, ring lattices, and all-to-all neuronal networks}
\author{M. A. Lopes}
\email{m.lopes@exeter.ac.uk}
\affiliation{Living Systems Institute, University of Exeter, Devon EX4, United Kingdom}
\affiliation{Centre for Biomedical Modelling and Analysis, University of Exeter, Devon EX4, United Kingdom}
\affiliation{EPSRC Centre for Predictive Modelling in Healthcare, University of Exeter, Devon EX4, United Kingdom}
\affiliation{Department of Physics $\&$ I3N, University of Aveiro, 3810-193 Aveiro, Portugal}
\author{A. V. Goltsev}
\affiliation{Department of Physics $\&$ I3N, University of Aveiro, 3810-193 Aveiro, Portugal}
\affiliation{A.F. Ioffe Physico-Technical Institue, 194021 St. Petersburg, Russia}

\begin{abstract}
Neuronal network dynamics depends on network structure. In this paper we study how network topology underpins the emergence of different dynamical behaviors in neuronal networks. In particular, we consider neuronal network dynamics on Erd\H{o}s--R\'enyi (ER) networks, regular random (RR) networks, ring lattices, and all-to-all networks. We solve analytically a neuronal network model with stochastic binary-state neurons in all the network topologies, except ring lattices. Given that apart from network structure, all four models are equivalent, this allows us to understand the role of network structure in neuronal network dynamics. Whilst ER and RR networks are characterized by similar phase diagrams, we find strikingly different phase diagrams in the all-to-all network. Neuronal network dynamics is not only different within certain parameter ranges, but it also undergoes different bifurcations (with a richer repertoire of bifurcations in ER and RR compared to all-to-all networks). This suggests that local heterogeneity in the ratio between excitation and inhibition plays a crucial role on emergent dynamics. Furthermore, we also observe one subtle discrepancy between ER and RR networks, namely ER networks undergo a neuronal activity jump at lower noise levels compared to RR networks, presumably due to the degree heterogeneity in ER networks that is absent in RR networks. Finally, a comparison between network oscillations in RR networks and ring lattices shows the importance of small-world properties in sustaining stable network oscillations.
\end{abstract}

\maketitle

\section{Introduction}
\label{introduction}
	
The brain is an enormous network of neurons connected by synapses. Neurons are dynamical systems whose dynamics depends on the interaction with other neurons. Understanding how network structure shapes emergent neuronal dynamics is of fundamental importance to unveil the workings of the brain. Modelling of neuronal networks has often considered neurons connected in all-to-all or random networks (see e.g. Refs. \cite{Brunel_1999, Brunel_2000, Koulakov_2002, Borgers_2003, Izhikevich_2003, Yuste_2015}). 

Many models in statistical physics, including the Ising, Potts, Kuramoto and other models, demonstrate the standard mean-field behavior in random networks, as in all-to-all networks, provided that the heterogeneity of the network is sufficiently weak, namely, when the second moment of the degree distribution is finite \cite{Dorogovtsev_2002b,Dorogovtsev_2008,Lopes_2016}. Additionally, the annealed network approximation by which an uncorrelated random network may be replaced by a weighted all-to-all network \cite{Giuraniuc_2006, Dorogovtsev_2008, Lopes_2016} further suggests that representing a random network with an all-to-all network may be an acceptable approximation. Regular random (RR) networks have also been used to obtain mean-field solutions which, depending on the applications, may be concordant with both random and all-to-all networks \cite{Dorogovtsev_2008} (note that in RR networks all nodes have the same number of connections, i.e. the same degree, in contrast to random networks where node  degree varies between nodes). However, such concordance depends on how nodes interact with each other. In the case of neuronal networks, it has long been understood that random and all-to-all networks underpin different emergent dynamics \cite{Wang_1995}, and careful considerations have been devoted to random networks \cite{Vreeswijk_1998, Brunel_2000, Kadmon_2015}. 

Herein we aim to better understand how network topology underpins the emergence of different dynamical behaviors in neuronal networks. We will consider the same neuronal model across Erd\H{o}s--R\'enyi (ER) networks, RR networks, ring lattices, and all-to-all networks, so that differences may only result from network topology. We focus on these four prototypical network structures because they enable us to reveal the role of key topological properties in the dynamics. On the other hand, ER, RR, and all-to-all networks are sufficiently simple to allow an analytical treatment. Although all-to-all networks (complete graphs), RR, and ER networks are all infinite dimensional systems \cite{Daqing_2011}, they have different topological and structural properties. In both ring lattices and RR networks considered in this paper, all nodes have the same degree, however in RR networks the nodes are randomly connected with other nodes, whereas in ring lattices they are connected to their closest neighbors. In ER networks, nodes are not only connected at random but also their degree varies across nodes. In all-to-all networks, all nodes are connected to all other nodes and therefore the distance between any two nodes is one, in contrast with RR and ER networks where the mean distance between any two nodes increases logarithmically with increasing size $N$, i.e., as $\log(N)$, which also differs from the power law dependence $N^{1/d}$ in any $d$--dimensional lattice, particularly in a ring lattice where the distance increases linearly with $N$. As a result the mean distance between any two nodes in random complex networks, including RR networks, is much smaller on average than in any $d$--dimensional lattice of the same size \cite{Watts_1998}. This small-word property enhances synchronization between interacting units in random complex networks \cite{Barahona_2002}. Additionally, clustering is large in all-to-all networks, whereas in RR and ER networks it is zero in the thermodynamic limit. Such differences may help us understand the role of network heterogeneity in emergent dynamics, namely whether synchronization is mostly promoted by clustering, or small-world properties.

\section{Model}
We consider the neuronal network model introduced in Refs.~\cite{Goltsev_2010,Lee_2014} and further studied in Refs.~\cite{Holstein_2013, Lopes_2014, Lopes_2017}. The network consists of $N$ neurons, $g_eN$ excitatory neurons, and $g_iN$ inhibitory neurons ($g_e+g_i=1$). Neurons can either be active and fire spike trains or be inactive and stay silent. Their state is a function of positive currents coming from presynaptic excitatory neurons and negative currents from presynaptic inhibitory neurons. Additionally, neurons are also stimulated by noise which accounts for both internal and external stochastic processes that may influence neuronal dynamics \cite{Faisal_2008}. The neurons act as stochastic integrators: they sum their input currents during an integration time $\tau$ and switch their dynamical state with probability $\mu_a\tau$ depending on whether the input is larger or smaller than a threshold $\Omega$. More specifically, an inactive excitatory (inhibitory) neuron becomes active with probability $\mu_e \tau $ ($\mu_i \tau $) if its total input current is larger than $\Omega$. Conversely, an active neuron becomes inactive with probability $\mu_a \tau $ if its total input current is smaller than $\Omega$ ($\mu_a=\mu_e$ for excitatory neurons, and $\mu_a=\mu_i$ for inhibitory neurons). $\mu_e^{-1}$ and $\mu_i^{-1}$ are the first-spike latencies of excitatory and inhibitory neurons, respectively. As we shall see, the ratio $\alpha=\mu_i/\mu_e$ plays an important role in the model by controlling the relative response times of excitatory and inhibitory neurons.

We define the fractions of active excitatory and inhibitory neurons at time $t$, $\rho_e(t)$ and $\rho_i(t)$, to characterise the neuronal network dynamics. We will refer to these fractions as activities. These activities follow the rate equations \cite{Goltsev_2010, Lee_2014}
\begin{equation}
\frac{\dot{\rho_a}}{\mu_a}=-\rho_a+\Psi_a(\rho_e,\rho_i),
\label{rho_eq}
\end{equation}
where $a=e,i$, $\dot{\rho}\equiv d\rho/dt$, and $\Psi_a(\rho_e,\rho_i)$ is the probability of a randomly chosen neuron to become active at time $t$. This function $\Psi_a$ encodes all information concerning single neuron dynamics, noise, and network structure. We will consider four network topologies: Erd\H{o}s--R\'enyi networks, regular random networks, random ring lattices, and all-to-all networks.

\subsection{Erd\H{o}s--R\'enyi network}
We have previously solved the model in the case where neurons are connected in a Erd\H{o}s--R\'enyi network \cite{Goltsev_2010, Lee_2014}. We found the heterogeneous mean-field function $\Psi_a(\rho_e,\rho_i)\equiv \Psi_{ER}(\rho_e,\rho_i)$,
\begin{eqnarray}
\Psi_{ER}(\rho_e,\rho_i)&=&\sum_{k,l,n=0}^\infty\Theta(J_ek+J_il+n-\Omega)   \nonumber \\
&\times& P_k(g_e\rho_ec) P_l(g_i\rho_ic)G(n,\langle n \rangle,\sigma).
\label{psi_rand}
\end{eqnarray}
The function considers a randomly chosen neuron that integrates $k$ spikes from excitatory presynaptic neurons, $l$ spikes from inhibitory presynaptic neurons, and $n$ spikes from noise. $J_e$ and $J_i$ are synaptic efficacies that weight these contributions ($J_e>0$ and $J_i<0$). $\Theta(x)$ is the Heaviside step function, $\Theta(x)=1$ if  $J_ek+J_il+n>\Omega$, otherwise $\Theta(x)=0$. The numbers of excitatory and inhibitory spikes, $k$ and $l$, follow a Poisson distribution, $P_n(\lambda)\equiv \lambda^n e^{-\lambda}/n!$, that accounts for the random structure \cite{Goltsev_2010}. The average number of spikes $\lambda$ is $g_a\rho_ac$, where $c$ is the mean in-degree, and it accounts for the average fraction of active presynaptic neurons in population $a$. The noise follows a Gaussian distribution $G(n,\langle n \rangle,\sigma)$ with mean $\langle n \rangle$ and variance $\sigma^2$ as in Refs.~\cite{Lee_2014, Lopes_2014, Lopes_2017}. For more details about the derivation of this function see Refs.~\cite{Goltsev_2010, Lee_2014}.

\subsection{Regular random network}
To study the role of topology in neuronal network dynamics, and particularly the role of randomness of the topology, we also consider neurons connected in a RR network. In this case, neurons are connected at random but the number of incoming (presynaptic) connections is constant and equal $c$. Thus, different neurons are connected to different numbers of excitatory and inhibitory neurons, though the total number of connections of every neuron is the same. The probability $p_1(n)$ that a randomly chosen neuron has $n$ excitatory and $c-n$ inhibitory presynaptic neighbors is
\begin{equation}
p_1(n)=\binom{c}{n}g_e^{n}g_i^{c-n},
\end{equation}
where $\binom{c}{n}$ is the binomial coefficient, $c!/(n!(c-n)!)$. Consequently, the probability $p_2(k,l)$ that a randomly chosen neuron receives $k$ spikes from active excitatory neurons and $l$ spikes from active inhibitory neurons during an integration time $\tau$ is
\begin{eqnarray}
p_2(k,l) &=& \sum_{n\ge k}^{c-l}p_1(n)\binom{n}{k}\rho_e^k(1-\rho_e)^{n-k} \nonumber \\
&\times& \binom{c-n}{l}\rho_i^l(1-\rho_i)^{c-n-l}.
\end{eqnarray}
Here we define that an active neuron fires one spike per integration time. This assumption provides qualitatively equivalent neuronal network dynamics when compared to lower or higher spiking rates in this model \cite{Goltsev_2010}. The probability $p_2(k,l)$ can be further simplified by using the binomial theorem,  
\begin{equation}
p_2(k,l) = c! \frac{(g_e\rho_e)^k}{k!}\frac{(g_i\rho_i)^l}{l!}\frac{(1-g_e\rho_e-g_i\rho_i)^{c-k-l}}{(c-k-l)!},
\end{equation}
and by introducing the Poisson distribution,
\begin{equation}
p_2(k,l) = \frac{c!e^c}{c^c} P_k(g_e\rho_ec) P_l(g_i\rho_ic)P_{c-l-k}(c[1-g_e\rho_e-g_i\rho_i]).
\label{eq-p2}
\end{equation}
Thus, one can show that the probability of a randomly chosen neuron to be active in the RR network is
\begin{eqnarray}
\Psi_{RR}(\rho_e,\rho_i)&=&\sum_{n=0}^\infty\sum_{k=0}^c\sum_{l=0}^{c-k}\Theta(J_ek+J_il+n-\Omega) \nonumber \\  
&\times& p_2(k,l)G(n,\langle n \rangle,\sigma),
\label{eq-psi_rr}
\end{eqnarray}
where we sum over all possible numbers of incoming spikes from noise ($n$), active excitatory presynaptic neighbors ($k$), and active inhibitory presynaptic neighbors ($l$). The Heaviside step function imposes that a neuron may only become active if $J_ek+J_il+n>\Omega$, $p_2(k,l)$ defines the probability of receiving $k$ and $l$ spikes from presynaptic neurons, and $G(n,\langle n \rangle,\sigma)$ is the probability of being excited by $n$ spikes from noise. By substituting Eq.~(\ref{eq-p2}) into Eq.~(\ref{eq-psi_rr}) and using Stirling's approximation, truncating the sum over $n$, and rearranging the sums, we obtain
\begin{eqnarray}
\Psi_{RR}(\rho_e,\rho_i)&\approx&\sqrt{2\pi c}\sum_{k=0}^c P_k(g_e\rho_ec)\sum_{l=0}^{c-k} P_l(g_i\rho_ic) \nonumber \\ 
&\times& P_{c-l-k}(c[1-g_e\rho_e-g_i\rho_i]) \\
&\times& \sum_{n=\Omega-J_ek-J_il}^{\langle n \rangle+3\sigma} G(n,\langle n \rangle,\sigma). \nonumber
\end{eqnarray}
Note that $\Psi_{RR}$ differs from $\Psi_{ER}$ in three aspects: (i) the coefficient $\sqrt{2\pi c}$; (ii) the sums over $k$ and $l$ are truncated (given that neurons may receive spikes from up to $c$ presynaptic neurons); and (iii) the function $P_{c-l-k}(c[1-g_e\rho_e-g_i\rho_i])$.

\subsection{Ring lattice}
To further understand the role of randomness in the topology in emerging network dynamics, we also consider ring lattices. In this case, each node on a ring with $N$ nodes is connected to all nodes placed at a distance smaller or equal to $c$. For simplicity, we consider all connections with the same direction, i.e. all connections coming from the left are in-connections whereas all connections to the right are out-connections. Finally, $Ng_e$ excitatory and $Ng_i$ inhibitory neurons are distributed at random over the $N$ nodes. (Undirected regular ring lattices were used in the seminal paper of Watts and Strogatz \cite{Watts_1998} to build small-world networks: small-world properties were obtained by randomly rewiring a fraction of all connections of the lattice.) Note that in the RR network, neurons are connected at random and consequently the mean distance between any two neurons increases as $\log(N)$, which is much smaller than the mean distance between two neurons in the ring lattice where the distance grows linearly with the system size $N$. However, the considered directed ring lattice has the same distribution of pre- and postsynaptic excitatory and inhibitory neurons as the RR network. As in RR networks, each neuron in a ring lattice is connected to a random number $n_e$ of excitatory and $n_i$ of inhibitory presynaptic neurons, whose sum $n_e+n_i$ is $c$. For this network topology we do not have an analytical solution and consequently we limited our analysis to simulations of large networks of size $N=10^4$ and $N=10^5$. We explain the algorithm to generate simulations below, in Sec.~\ref{parameters_simulations}.

\subsection{All-to-all network}
\label{net_all}
Finally, we further consider neurons connected in an all-to-all network, where every neuron is topologically equivalent to all other neurons. Whilst from the ER to the RR network we removed randomness from the topology but kept randomness in the distribution of excitatory and inhibitory neurons across the network, from the regular to the all-to-all network we are also removing this heterogeneity: all neurons are connected to the same number of excitatory and inhibitory neurons. In this case, every neuron receives spikes from all other active neurons in the network,
\begin{eqnarray}		
J_ek &=& J_eg_e\rho_e(N-1) =  \tilde{J}_eg_e\rho_e \nonumber \\
J_il &=& J_ig_i\rho_i(N-1) =  \tilde{J}_ig_i\rho_i,
\end{eqnarray}
where we use the standard normalisations, $J_e\to\tilde{J}_e/(N-1)$ and $J_i\to\tilde{J}_i/(N-1)$. Note that these normalisations imply that both the noise intensity $n$ and threshold $\Omega$ must be rescaled. Given that, in the case of ER networks, the input current in Eq.~(\ref{psi_rand}) is proportional to the mean in-degree $c$, for the sake of comparison we define $\eta=n/c$, $\omega=\Omega/c$, and consequently $\langle \eta \rangle = \langle n \rangle/c$, and $\tilde{\sigma} = \sigma/c$. We thus find the $\Psi_{all}$ function for an all-to-all network,
\begin{equation}
\Psi_{all}(\rho_e,\rho_i)=\sum_{\eta=0}^\infty \Theta(\tilde{J}_eg_e \rho_e+\tilde{J}_ig_i \rho_i+\eta-\omega)G(\eta,\langle \eta \rangle,\tilde{\sigma}).
\label{psi2_eq}
\end{equation}
As above, we consider Gaussian noise and therefore $\Psi_{all}(\rho_e,\rho_i)$ can be written as
\begin{equation}
\Psi_{all}(\rho_e,\rho_i)=\Phi\Big(\frac{\tilde{J}_eg_e\rho_e+\tilde{J}_ig_i\rho_i+\langle \eta \rangle-\omega}{\tilde{\sigma}}\Big).
\end{equation}
where $\Phi(x)$ is the cumulative distribution function of the standard normal distribution \cite{Abramowitz_1970},
\begin{equation}
\Phi(x)=\frac{1}{\sqrt{2\pi}}\int_{-\infty}^x e^{-x^2/2}dt.
\end{equation}
Thus, the neuronal network dynamics in all-to-all networks are governed by the following rate equations
\begin{equation}
\frac{\dot{\rho_a}}{\mu_a} =-\rho_a+\Phi\Big(\frac{\tilde{J}_eg_e\rho_e+\tilde{J}_ig_i\rho_i+\langle \eta \rangle-\omega}{\tilde{\sigma}}\Big),
\label{drho_all}
\end{equation}
where $a=e,i$.

\subsection{Parameters and numerical simulations}
\label{parameters_simulations}
We consider the following model parameters. In ER networks, RR networks, and ring lattices we use the mean in-degree $c=1000$, the threshold $\Omega=30$, the integration time $\tau=0.1\mu_e^{-1}$, the synaptic efficacies $J_e=1$ and $J_i=-3$, and the noise variance $\sigma^2=10$. These parameters have been discussed and justified elsewhere \cite{Goltsev_2010, Lee_2014, Lopes_2014}. Analogously, in all-to-all networks we use $\omega=\Omega/c=0.03$, the integration time $\tau=0.1\mu_e^{-1}$, $\tilde{J}_e=1$, $\tilde{J}_i=-3$, and $\tilde{\sigma}^{2}=(\sigma/c)^2=10^{-5}$. The algorithm employed in our numerical simulations was explained in \cite{Goltsev_2010, Lee_2014}. Briefly, we constructed directed ER networks by connecting neurons with probability $c/N$, whereas to obtain directed regular networks we built regular ring lattices and rewired links randomly while preserving the degree distribution using the Maslov-Sneppen rewiring algorithm \cite{Maslov_2002}. Ring lattices were obtained by connecting each neuron to its closest $c$ pre-synaptic neighbors. Finally, all-to-all networks were built by connecting all nodes to all other nodes except themselves. In all network topologies, nodes were randomly assigned as being excitatory or inhibitory, such that the total number of excitatory and inhibitory neurons were $g_eN$ and $g_iN$, respectively. Time was discretized into intervals $\Delta t = \tau$. We initialized our simulations with all neurons inactive. We then evaluated at each time step whether the total input to each node was higher or lower than the threshold $\Omega$. The total input accounted for all presynaptic active neurons and gaussian noise as described above. Subsequently, the state of all neurons was updated in parallel at every time step depending on the individual total inputs following the rules stated above.

\section{Steady states}
To characterise and compare the neuronal dynamics across different network topologies, we first find the steady states in each network. In this section we focus on ER, regular, and all-to-all networks, leaving out ring lattices, for which we do not have an analytical solution. The neuronal networks reach a steady state when $d\rho_a/dt=0$. In all three networks, steady excitatory activity is equal to steady inhibitory activity, $\rho_e=\rho_i\equiv\rho$. In ER and RR networks, we find the steady state equations 
\begin{equation}
\rho=\Psi_{ER}(\rho,\rho)
\label{rho_rand}
\end{equation}
and
\begin{equation}
\rho=\Psi_{RR}(\rho,\rho),
\label{rho_reg}
\end{equation}
respectively. Similarly, we find the steady state equation in all-to-all networks 
\begin{equation}
\rho=\Phi\Big(\frac{\tilde{J}_eg_e\rho+\tilde{J}_ig_i\rho+\langle \eta \rangle-\omega}{\tilde{\sigma}}\Big).
\label{rho_all}
\end{equation} 
Solutions of these equations were obtained by solving numerically the right-hand side for $800$ values of $\rho$ in the range $[0,1]$ and then finding the graphical intersection with $\rho$.

\begin{figure}
\includegraphics[width=0.49\textwidth]{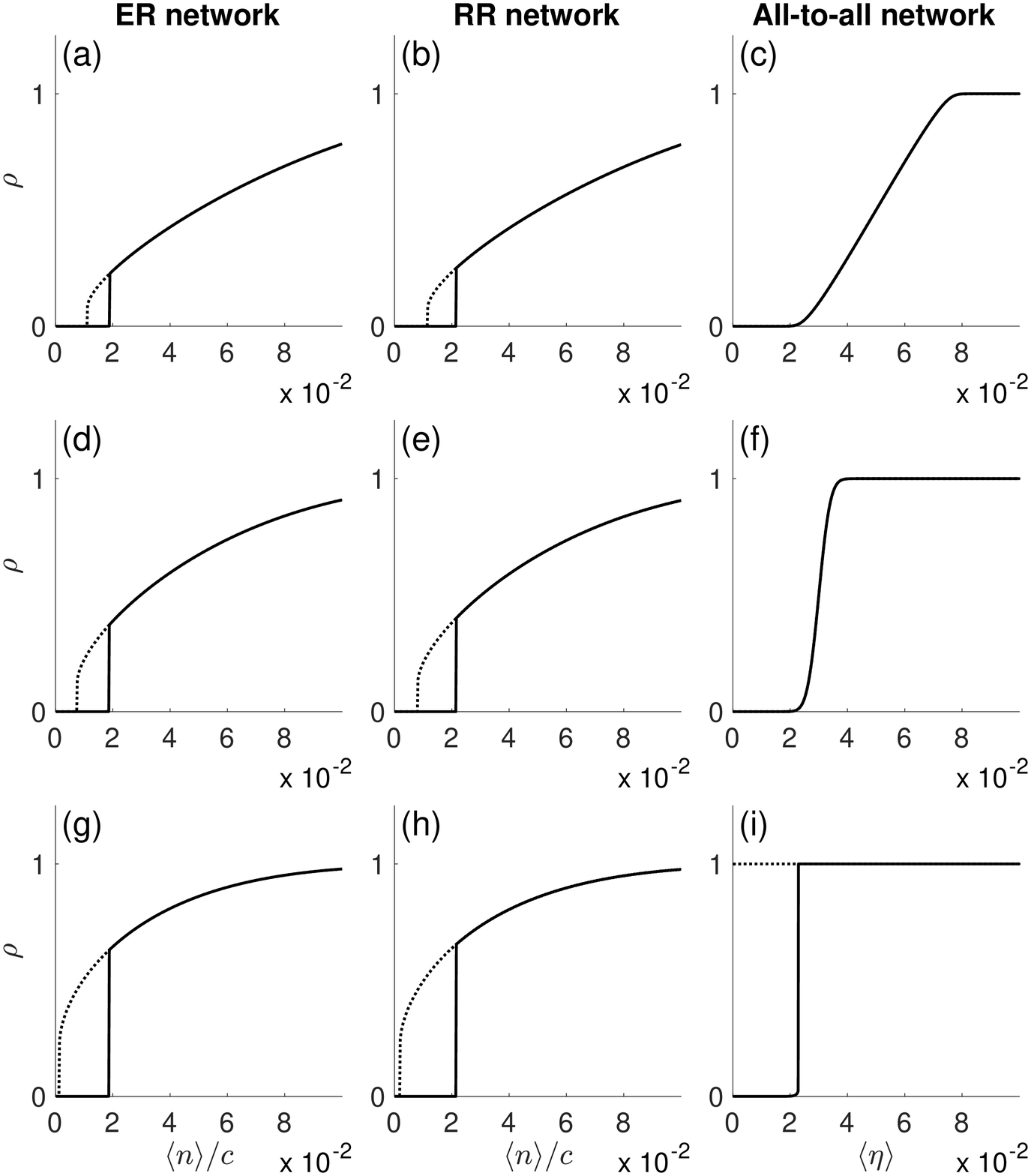}
\caption{Steady state neuronal activity $\rho$ as a function of the level of noise $\langle n \rangle /c$ and $\langle \eta \rangle$ in ER (left column), RR (middle column) and all-to-all networks (right column). These steady states are the result of the numerical integration of Eqs.~(\ref{rho_rand})--(\ref{rho_all}). Each row corresponds to networks with different fractions of excitatory neurons: (a)-(c) $g_e=0.74$, (d)-(f) $g_e=0.75$, and (g)-(i) $g_e=0.76$. The dashed lines represent upper metastable states in bistability regions where $\rho$ may take low or high activity values depending on the initial conditions. \label{fig_steady}}
\end{figure}

Figure~\ref{fig_steady} shows the steady states $\rho$ as a function of the noise intensity in networks with different fractions of excitatory neurons $g_e$. The noise has an excitatory effect on neurons and as a result $\rho$ grows with increasing noise. We also find a strong dependence of $\rho$ on $g_e$. Note that at $g_e=0.75$ the network is balanced, i.e. $g_eJ_e=g_i|J_i|$, and therefore the quantity $J_eg_e\rho_e+J_ig_i\rho_i=(J_eg_e+J_ig_i)\rho$ is zero at the steady states, whilst it is negative at $g_e=0.74$ and positive at $g_e=0.76$. We observe that larger fractions of $g_e$ are responsible for more pronounced increases of $\rho$ as a function of noise. However, although we find a bistability region bounded by activity jumps in both ER and RR networks at intermediate noise levels (panels in the left and middle columns), all-to-all networks show no bistability when $g_e=0.74$ and $g_e=0.75$, and instead $\rho$ grows gradually with increasing noise $\langle \eta \rangle$. The steepness of $\rho$ as a function of $\langle \eta \rangle$ gets higher with increasing $g_e$, and a bistability region emerges when the steepness becomes infinite. Panel (i) further shows that the bistability region appears in all-to-all networks only at $g_e > 0.75$, bounded by $\langle \eta \rangle =0$. In contrast, ER and RR networks display a bistability region at $g_e$ both above and below $0.75$, and at $g_e=0.76$ the region is bounded by a bifurcation point $\langle n \rangle >0$. Finally, we observe that although the steady states in ER and RR networks are very similar, the bifurcation point at which there is an activity jump is slightly higher in RR networks compared to ER networks. We interpret this difference as a consequence of a lower heterogeneity in RR networks compared to ER networks. In ER networks there is a higher chance of finding neurons with higher number of presynaptic excitatory neurons compared to RR networks, given that in RR networks neurons have at most $c$ excitatory presynaptic neurons. A higher number of 'hyper-excitable' neurons may enable ER networks to jump to higher activities at lower levels of noise. 

\section{Phase diagrams and dynamics}
To further characterise the neuronal dynamics, we study the local stability of the fixed points determined by Eqs.~(\ref{rho_rand})--(\ref{rho_all}) \cite{Strogatz_1994, Lee_2014}. This stability is determined by the eigenvalues of the Jacobian of Eqs.~(\ref{rho_eq}),
\begin{equation}
\widehat{J}(\rho) =
\begin{pmatrix}
-1 +\partial \Psi/\partial \rho_e & \partial \Psi/\partial \rho_i \\
\alpha \partial \Psi/\partial \rho_e & -\alpha +\alpha \partial \Psi/\partial \rho_i
\end{pmatrix},
\label{Jacobian}
\end{equation}
at the fixed points $\rho$. In the case of the all-to-all network, the Jacobian of the dynamical system described by Eqs.~(\ref{drho_all}) is 
\begin{equation}
\widehat{J}(\rho) =
\begin{pmatrix}
-1+ \tilde{J}_eg_eG(x) & \tilde{J}_ig_iG(x) \\
\alpha \tilde{J}_eg_eG(x) & -\alpha+ \alpha \tilde{J}_ig_iG(x)
\end{pmatrix}.
\end{equation}
where $G(x)$ is the Gaussian distribution with zero mean and standard deviation $\tilde{\sigma}$,
\begin{equation}
G(x)=\frac{1}{\sqrt{2\pi \tilde{\sigma}^{2}}}e^{-\frac{x^2}{2\tilde{\sigma}^{2}}},
\end{equation}
and $x=\tilde{J}_eg_e\rho+\tilde{J}_ig_i\rho+\langle \eta \rangle-\omega$.

The eigenvalues of the Jacobian matrices are given by 
\begin{equation}
\lambda_{\pm}=-\frac{1}{2}(J_{11}+J_{22})\pm \frac{1}{2}\sqrt{(J_{11}-J_{22})^2+4J_{12}J_{21}},
\label{eigenv}
\end{equation}
where $J_{ij}$ are the entries of the Jacobian. 

To find phase boundaries separating different dynamical behaviors in ER, RR, and all-to-all networks, we solve the conditions
\begin{equation}
\text{Re}\{\lambda_\pm\} =0,
\label{eig_real}
\end{equation}
and
\begin{equation}
\text{Im}\{\lambda_\pm\}=0.
\label{eig_imag}
\end{equation}
The fact that the steady state equations (\ref{rho_rand})--(\ref{rho_all}) do not depend on $\alpha$ allows us to find $\alpha$ as a function of the level of noise at which these conditions, Eqs.~(\ref{eig_real}) and (\ref{eig_imag}), are satisfied. Additionally, we solve the equation
\begin{equation} 
\frac{\partial \Psi(\rho,\rho)}{\partial \rho}=1
\label{saddle-node}
\end{equation}
which determines the level of noise at which the neuronal activity jumps observed in Fig.~\ref{fig_steady} take place. This condition actually defines the coalescence or emergence of fixed points, i.e. the bifurcation point at which the steady state equations (\ref{rho_rand})--(\ref{rho_all}) transit from one solution to three, or vice-versa \cite{Lee_2014}. We have previously demonstrated that the jumps correspond to saddle-node bifurcations \cite{Lee_2014}. 

\begin{figure}
\includegraphics[width=0.49\textwidth]{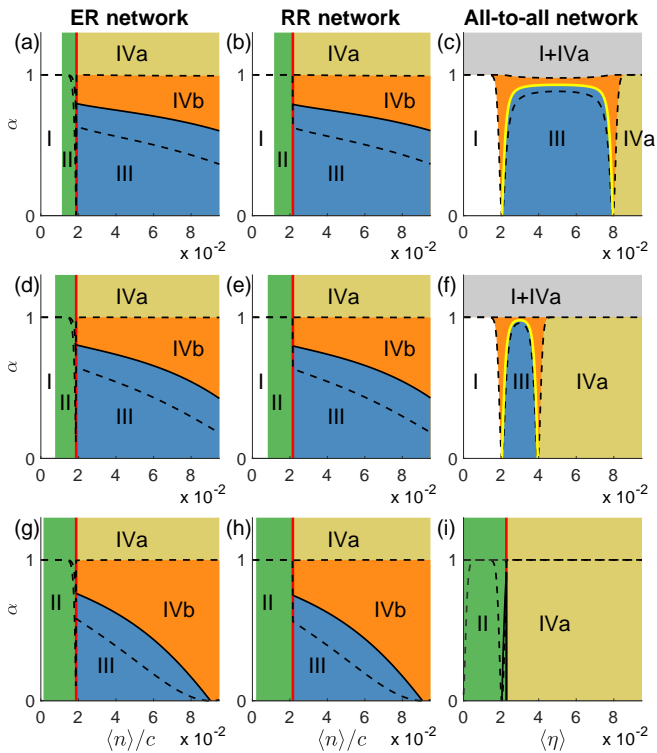}
\caption{Noise--$\alpha$ planes of the phase diagram of the neuronal network models. Left, middle, and right columns correspond respectively to ER, RR, and all-to-all networks. Each row represents networks with different fractions of excitatory neurons: (a)-(c) $g_e=0.74$, (d)-(f) $g_e=0.75$, and (g)-(i) $g_e=0.76$. There are four regions of activity: (I) low neuronal activity; (II) bistability region; (III) neuronal network oscillations; and (IV) high neuronal activity with (a) exponential relaxation and (b) damped oscillations. All-to-all networks have a region I+IVa which contains a continuum from low to high activity as a function of increasing noise intensity $\eta$. The black and yellow solid lines are the numerical solutions of Eq.~(\ref{eig_real}), whereas the black dashed lines are the numerical solutions of Eq.~(\ref{eig_imag}). The black solid lines correspond to supercritical Hopf bifurcations and the yellow solid lines represent subcritical Hopf bifurcations. The red lines correspond to saddle-node bifurcations determined by Eq.~(\ref{saddle-node}). 
\label{fig_phaseDiagram}}
\end{figure}

Figure~\ref{fig_phaseDiagram} shows the numerical solutions of Eqs.~(\ref{eig_real})--(\ref{saddle-node}) in noise--$\alpha$ planes at different fractions of excitatory neurons $g_e$ for the three network topologies. We identify four regions of neuronal activity: in region I the activity relaxes exponentially to a low activity state; region II is a bistability region where the lower and upper metastable states may be stable or unstable (see Ref.~\cite{Lee_2014} for more details); region III corresponds to sustained network oscillations; and in region IVa and IVb the activity relaxes exponentially and in the form of damped oscillations to a high activity state, respectively. Note that in all-to-all networks (at $g_e=0.74, 0.75$), the absence of a saddle-node bifurcation enables regions I and IVa to form a continuum from low to high activity at sufficiently high $\alpha$ (region I+IVa in Fig.~\ref{fig_phaseDiagram}(c) and (f)). We observe that as we increase the fraction of excitatory neurons $g_e$, the region of neuronal network oscillations shrinks in the three network topologies. At $g_e=0.76$, the all-to-all network no longer displays network oscillations in striking contrast with ER and RR networks which present a large area in parameter space with oscillations. Furthermore, we find that whilst region III in Fig.~\ref{fig_phaseDiagram}(a-b),(d-e) and (g-h) is bounded on the left (at a low noise intensity) by a saddle-node on invariant circle (SNIC) bifurcation and, on the right (at a high noise intensity), by a supercritical Hopf bifurcation in ER and RR networks, instead oscillations in all-to-all networks emerge only due to a subcritical Hopf bifurcation. Thus, in ER and RR networks oscillations emerge above the bifurcation point $n_{c1}$ of the SNIC bifurcation with a finite amplitude but a small frequency proportional to $(\langle n \rangle -n_{c1})^{1/2}$, whereas close to the supercritical Hopf bifurcation, the oscillations have a finite frequency with an amplitude that decreases proportionally to $(n_{c2}-\langle n \rangle )^{1/2}$ as we approach the bifurcation point $n_{c2}$. In contrast, in all-to-all networks oscillations emerge with both finite amplitude and frequency. In this case, however, there is a narrow parameter range with hysteresis, where the all-to-all network displays either damped oscillations or network oscillations depending on the initial conditions (this region is not represented in Fig.~\ref{fig_phaseDiagram}). Again, the only clear difference between ER and RR networks is the level of noise at which the SNIC bifurcation takes place: at a lower level of noise in ER networks compared to RR networks. 

\begin{figure}
\includegraphics[width=0.49\textwidth]{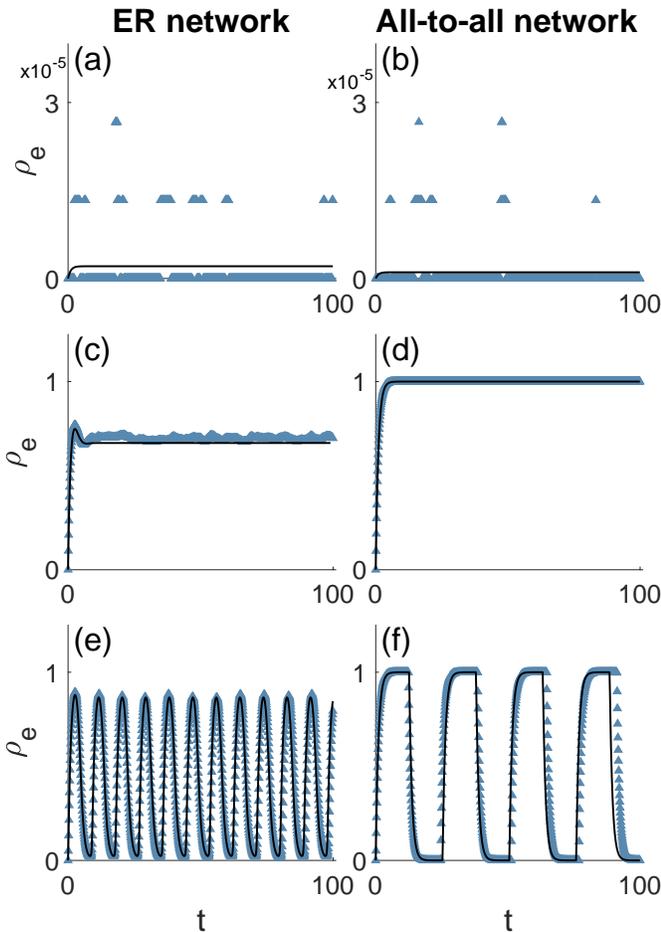}
\caption{Excitatory activity $\rho_e$ as a function of time in ER (left column), and all-to-all networks (right column). Panels (a)-(b) display low activity from region I in Fig.~\ref{fig_phaseDiagram}, $(\langle n \rangle/c,\alpha)=(\langle \eta \rangle,\alpha)=(0.015,0.7)$; panels (c)-(d) represent high activity from region IVb in ER networks and IVa in all-to-all networks, $(\langle n \rangle/c,\alpha)=(\langle \eta \rangle,\alpha)=(0.05,0.9)$; and panels (e)-(f) show network oscillations from region III, $(\langle n \rangle/c,\alpha)=(\langle \eta \rangle,\alpha)=(0.03,0.7)$. The black lines are the numerical solution of Eqs.~(\ref{rho_eq}) for each network topology, and the blue triangles represent numerical simulations of the model (number of neurons $N=10^{5}$). We used a fraction of excitatory neurons $g_e=0.75$.
 \label{fig_activity}}
\end{figure}

\begin{figure}
\includegraphics[width=0.49\textwidth]{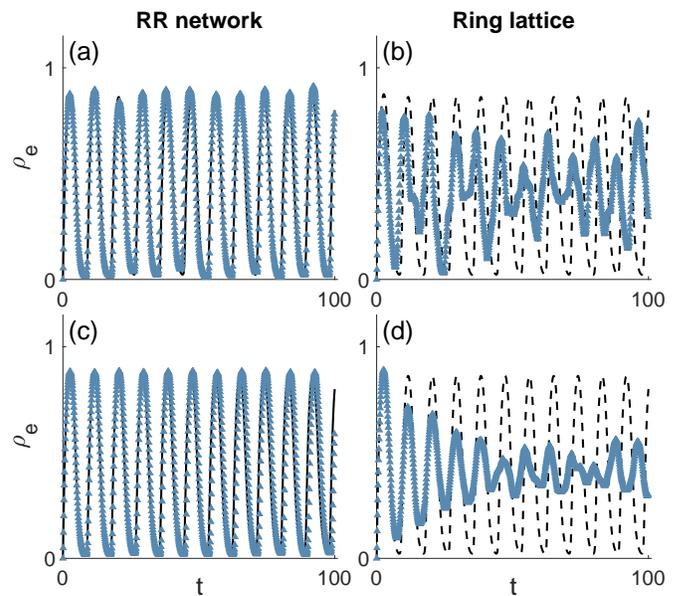}
\caption{Network oscillations in RR networks (left column), and in ring lattices (right column). As in Fig.~\ref{fig_activity}, the blue triangles represent numerical simulations of the model in finite networks. The black lines are the numerical solution of Eqs.~(\ref{rho_eq}) for RR networks. The same numerical solutions are plotted as dashed lines in the right column for comparison with the simulations in finite ring lattices. Panels (a) and (b) correspond to networks with size $N=10^4$, whereas panels (c) and (d) display oscillations in networks with size $N=10^5$. We used the same parameters in all the panels: $(\langle n \rangle/c,\alpha)=(0.03,0.7)$, and a fraction of excitatory neurons $g_e=0.75$.
 \label{fig_oscillations}}
\end{figure}

Figure~\ref{fig_activity} displays representative neuronal network activity in three of the regions identified in Fig.~\ref{fig_phaseDiagram}. We chose equivalent parameters in the three networks corresponding to comparable regions of the phase diagrams, but decided to only show here the activity in ER and all-to-all networks because RR networks displayed activities almost indistinguishable from the activities in ER networks. As expected taking into account Fig.~\ref{fig_steady}, the steady states are quantitatively different across the three networks, though qualitatively similar. However, we observe that network oscillations in all-to-all networks have a different shape compared to oscillations in ER and RR networks, where they are almost equivalent (compare Fig.~\ref{fig_activity}(e) and Fig.~\ref{fig_oscillations}(c)). Figure~\ref{fig_activity} also shows the result of simulations using networks comprising $10^{5}$ neurons. Note that in the low activity state, panels (a) and (b), the activity $\rho_e$ is smaller than $1 / N$ hence most neurons are silent most of the time in the simulations except for occasional random firings. For comparison, we observed the steady states $\rho_e=(2.08\times10^{-6}, 0, 1.05\times10^{-6})$ from the numerical integration of Eqs.~(\ref{rho_eq}) for ER, RR, and all-to-all networks, respectively, which are in good agreement with the average activities from simulations, $\langle \rho_e \rangle = (1.92\times10^{-6},4.27\times10^{-7},9.60\times10^{-7})$. In RR networks, random fluctuations can also sporadically activate neurons, but at a smaller rate compared to ER and all-to-all networks. In the high activity state, whilst neuronal activity fluctuates in ER (and RR) networks close to the steady states (see panel (c)), it does not in all-to-all networks (see panel (d)). In all three networks, we observe a good agreement with respect to network oscillations when comparing finite neuronal networks and the numerical integration of Eqs.~(\ref{rho_eq}) (corresponding to the infinite size limit).

We further compared simulations of RR networks with ring lattices. Note that a ring lattice is in fact a particular network realization of a RR network, where all neurons happen to be connected to their closest presynaptic neighbors. The two networks have the same in- and out-degree distributions of excitatory and inhibitory neurons. From this perspective, one could expect similar dynamics. However, as mentioned in the Introduction, the ring lattice is a one-dimensional system, whereas RR networks are infinite dimensional systems \cite{Daqing_2011}. We performed simulations of both neuronal network dynamics and indeed observed similar activity patterns, except for the region of network oscillations. In Fig.~\ref{fig_oscillations} we show network oscillations in RR networks and ring lattices. We observe that oscillations present lower and irregular amplitude in finite ring lattices in striking contrast with network oscillations in finite RR networks. Furthermore, whilst oscillations in finite RR networks approach the analytical solution as we increase the number of neurons $N$ (see the amplitude), instead they remain irregular and with lower amplitude in finite ring lattices. Nevertheless, the frequency of the oscillations is similar across RR and ring lattices.

\section{Discussion and Conclusions}
In this paper, we compared neuronal network dynamics across Erd\H{o}s--R\'enyi networks, regular random networks, ring lattices, and all-to-all networks using the same neuronal model in the three topologies. The considered model comprised stochastic binary-state excitatory and inhibitory neurons interacting together in a network \cite{Goltsev_2010, Lee_2014, Lopes_2014, Lopes_2017}. We found that network structure has a strong impact on the observed dynamics and bifurcation diagram. In particular, all-to-all networks underpin strikingly different dynamics compared to ER and RR networks in certain parameter ranges. On the other hand, ER and RR networks display very similar dynamics. This suggests that the randomness in the total number of presynaptic excitatory and inhibitory connections does not play a major role in the dynamics of these networks, provided that neurons are connected at random. In other words, local heterogeneity in the ratio of connections to presynaptic excitatory and inhibitory neurons may play a crucial role in neuronal network dynamics, particularly in neuronal oscillations and critical phenomena in the vicinity of bifurcations, whereas heterogeneity in the total number of presynaptic connections seems to be less relevant when we compare ER and RR networks. Furthermore, we observed that despite similarities between finite RR networks and finite ring lattices (they have the same pre- and postsynaptic degree distribution), network oscillations are fundamentally different in the two networks, a difference that becomes apparent as we increase the system size. 

Our results in Fig.~\ref{fig_steady} show that for balanced ($g_eJ_e=g_i|J_i|$) and slightly unbalanced networks towards inhibition ($g_eJ_e\lesssim g_i|J_i|$) there is bistability in ER and RR networks but not in all-to-all networks. At a fraction of excitatory neurons $g_e=0.76$ we found bistability in all three networks. However, the upper metastable state in ER and RR networks comprises about half the neuronal population, whereas the equivalent state in all-to-all networks involves the whole network. Such differences may help deciding whether a ER or an all-to-all network may be more appropriate to model, for example, neuronal cultures \cite{Orlandi_2013}. Interestingly, whilst we observe that the activity jump occurs at slightly higher noise levels in RR networks compared to ER networks, when we do observe a jump also in the all-to-all network, it occurs at a level of noise comparable to the one observed in the RR networks (but slightly larger). This supports our interpretation that ER networks may jump to the higher metastable state at lower levels of noise compared to both RR and all-to-all networks due to the existence of 'hyper-excitable' neurons (i.e. neurons with a higher imbalance in their excitatory and inhibitory presynaptic neighbors). Such neurons may also exist in RR networks, but their imbalance is bounded by the average in-degree.  

We found that fixed points characterized by complete activation of the network ($\rho\approx1$) are incompatible with oscillations in ER, RR, and all-to-all networks. Larger fractions of excitatory neurons $g_e$ in any of these networks lead to higher activities and consequently we observe that the region of network oscillations shrinks as we increase $g_e$. Interestingly, when we observed a region of oscillations in the three network structures [see Fig.~\ref{fig_phaseDiagram}(a)-(f)], this region appears to be symmetrical with regard to the level of noise in all-to-all networks, but not in ER and RR networks. More importantly, oscillations may emerge due to a SNIC bifurcation or a supercritical Hopf bifurcation in ER and RR networks, whereas in all-to-all networks the oscillatory regime is only bounded by a subcritical Hopf bifurcation, accompanied by hysteresis. Thus, whilst oscillations in ER and RR networks may have low frequency (and high-amplitude) close to the SNIC bifurcation, or low amplitude (and high-frequency) close to the supercritical Hopf bifurcation, in all-to-all networks oscillations have always finite amplitude and frequency. Although results in Fig.~\ref{fig_phaseDiagram} may seem to suggest that network oscillations vanish in all-to-all networks when the saddle-node bifurcation emerges, that is not actually the case. Further numerical analysis revealed that there is a narrow region of parameters at which the saddle-node bifurcation coexists with network oscillations in all-to-all networks, however the region of network oscillations remains bounded only by the Hopf bifurcation (results not presented here). 

We also demonstrated that even for parameters at which the three networks could be expected to be in similar dynamical regimes, we found some differences (see Fig.~\ref{fig_activity}). Whilst we found irregular fluctuations around a high activity state in ER and RR networks, we observed stable full network activation in all-to-all networks. Additionally, network oscillations also presented distinctive shapes in ER and RR networks compared to all-to-all networks. We further compared network oscillations in finite RR networks and finite ring lattices. Whilst oscillations were stable in RR networks and approached the analytical solution as we increased the network size $N$, the oscillations in the ring lattices were irregular and the amplitude seemed to decrease with increasing $N$. The main difference between a ring lattice and a RR network is the lack of small-world properties in the ring lattice, which restrains synchronization across the network \cite{Barahona_2002}. As the size of the network increases, the mean distance between nodes increases linearly with $N$, and therefore the communication between neurons is hindered. In fact, it is well-known in statistical physics that any interaction model in a one-dimensional system with short-range interactions cannot undergo a phase transition since fluctuations must destroy any long-ranged order in one dimensional systems at large times \cite{Stanley_1987}. Thus, network oscillations should not emerge in infinite ring lattices. Nevertheless, short-ranged correlations exist and the correlation length can be large, which can support the irregular oscillations observed in Fig.~\ref{fig_oscillations}. Interestingly, one can still observe dynamical similarities between oscillations in finite RR networks and ring lattices (though with strong fluctuations). We interpret the temporal behavior in ring lattices as a flickering dynamical behavior of the one found in RR networks.  

Based on these results, we would like to stress how profoundly network structure can influence network dynamics, particularly the differences between ER networks, ring lattices, and all-to-all networks. Note that ER and all-to-all networks are actually opposite ends in regard to clustering. The clustering coefficient of an undirected ER network is $c/N$, which tends to zero in the infinite-size limit \cite{Dorogovtsev_2002}. In contrast, the coefficient is $1$ in all-to-all networks. In undirected ring lattices, the clustering coefficient is also large: it is equal to $3(c-1)/[2(2c-1)]$ which tends to $3/4$ at $c\gg1$. Note that the clustering coefficient characterizes the occurrence of triplets in a network \cite{Dorogovtsev_2008}. Thus, whilst triplets may be neglected in ER networks, they may not in all-to-all networks and ring lattices. In our neuronal network there are many different triads since the network is directed and there are two types of nodes (excitatory and inhibitory neurons), which makes it difficult to predict how these motifs may influence the dynamics. Small-world properties and particularly large clustering coefficients have been observed in both large-scale brain networks \cite{Sporns_2004} and in neocortical microcircuitry \cite{Gal_2017}. At smaller scales, neurons are connected on average to about $10^4$ other neurons in the cortex \cite{Kandel_2000}, while packed in minicolumns \cite{Mountcastle_2003}, thus likely organised in dense clustered networks. Such high clustering promotes the emergence of rich dynamical patterns, as a recent study in networks of rat cortical neurons {\it in vitro} has shown \cite{Okujeni_2017}. Here we suggest that such rich dynamical behaviors may also be supported by local heterogeneities in excitation and inhibition across the network. Additionally, our results in ring lattices further support the importance of small-world properties in the emergence of synchronization \cite{Barahona_2002}.

\section{Acknowledgements}
This work was partially supported by FET IP Project MULTIPLEX 317532. A.V.G. is grateful to LA I3N for Grant No. PEST UID/CTM/50025/2013. M.A.L. acknowledges the financial support of the Medical Research Council (MRC) via grant MR/K013998/1. M.A.L. further acknowledges funding from Epilepsy Research UK via grant P1505.


\bibliography{Lopes_bib}

\begin{thebibliography}{33}%
\makeatletter
\providecommand \@ifxundefined [1]{%
 \@ifx{#1\undefined}
}%
\providecommand \@ifnum [1]{%
 \ifnum #1\expandafter \@firstoftwo
 \else \expandafter \@secondoftwo
 \fi
}%
\providecommand \@ifx [1]{%
 \ifx #1\expandafter \@firstoftwo
 \else \expandafter \@secondoftwo
 \fi
}%
\providecommand \natexlab [1]{#1}%
\providecommand \enquote  [1]{``#1''}%
\providecommand \bibnamefont  [1]{#1}%
\providecommand \bibfnamefont [1]{#1}%
\providecommand \citenamefont [1]{#1}%
\providecommand \href@noop [0]{\@secondoftwo}%
\providecommand \href [0]{\begingroup \@sanitize@url \@href}%
\providecommand \@href[1]{\@@startlink{#1}\@@href}%
\providecommand \@@href[1]{\endgroup#1\@@endlink}%
\providecommand \@sanitize@url [0]{\catcode `\\12\catcode `\$12\catcode
  `\&12\catcode `\#12\catcode `\^12\catcode `\_12\catcode `\%12\relax}%
\providecommand \@@startlink[1]{}%
\providecommand \@@endlink[0]{}%
\providecommand \url  [0]{\begingroup\@sanitize@url \@url }%
\providecommand \@url [1]{\endgroup\@href {#1}{\urlprefix }}%
\providecommand \urlprefix  [0]{URL }%
\providecommand \Eprint [0]{\href }%
\providecommand \doibase [0]{http://dx.doi.org/}%
\providecommand \selectlanguage [0]{\@gobble}%
\providecommand \bibinfo  [0]{\@secondoftwo}%
\providecommand \bibfield  [0]{\@secondoftwo}%
\providecommand \translation [1]{[#1]}%
\providecommand \BibitemOpen [0]{}%
\providecommand \bibitemStop [0]{}%
\providecommand \bibitemNoStop [0]{.\EOS\space}%
\providecommand \EOS [0]{\spacefactor3000\relax}%
\providecommand \BibitemShut  [1]{\csname bibitem#1\endcsname}%
\let\auto@bib@innerbib\@empty
\bibitem [{\citenamefont {Brunel}\ and\ \citenamefont
  {Hakim}(1999)}]{Brunel_1999}%
  \BibitemOpen
  \bibfield  {author} {\bibinfo {author} {\bibfnamefont {N.}~\bibnamefont
  {Brunel}}\ and\ \bibinfo {author} {\bibfnamefont {V.}~\bibnamefont {Hakim}},\
  }\href@noop {} {\bibfield  {journal} {\bibinfo  {journal} {Neural Comput.}\
  }\textbf {\bibinfo {volume} {11}},\ \bibinfo {pages} {1621} (\bibinfo {year}
  {1999})}\BibitemShut {NoStop}%
\bibitem [{\citenamefont {Brunel}(2000)}]{Brunel_2000}%
  \BibitemOpen
  \bibfield  {author} {\bibinfo {author} {\bibfnamefont {N.}~\bibnamefont
  {Brunel}},\ }\href {http://www.ncbi.nlm.nih.gov/pubmed/10809012} {\bibfield
  {journal} {\bibinfo  {journal} {J. Comput. Neurosci.}\ }\textbf {\bibinfo
  {volume} {8}},\ \bibinfo {pages} {183} (\bibinfo {year} {2000})}\BibitemShut
  {NoStop}%
\bibitem [{\citenamefont {Koulakov}\ \emph {et~al.}(2002)\citenamefont
  {Koulakov}, \citenamefont {Raghavachari}, \citenamefont {Kepecs},\ and\
  \citenamefont {Lisman}}]{Koulakov_2002}%
  \BibitemOpen
  \bibfield  {author} {\bibinfo {author} {\bibfnamefont {A.~A.}\ \bibnamefont
  {Koulakov}}, \bibinfo {author} {\bibfnamefont {S.}~\bibnamefont
  {Raghavachari}}, \bibinfo {author} {\bibfnamefont {A.}~\bibnamefont
  {Kepecs}}, \ and\ \bibinfo {author} {\bibfnamefont {J.~E.}\ \bibnamefont
  {Lisman}},\ }\href@noop {} {\bibfield  {journal} {\bibinfo  {journal} {Nat.
  Neurosci.}\ }\textbf {\bibinfo {volume} {5}},\ \bibinfo {pages} {775}
  (\bibinfo {year} {2002})}\BibitemShut {NoStop}%
\bibitem [{\citenamefont {B{\"o}rgers}\ and\ \citenamefont
  {Kopell}(2003)}]{Borgers_2003}%
  \BibitemOpen
  \bibfield  {author} {\bibinfo {author} {\bibfnamefont {C.}~\bibnamefont
  {B{\"o}rgers}}\ and\ \bibinfo {author} {\bibfnamefont {N.}~\bibnamefont
  {Kopell}},\ }\href@noop {} {\bibfield  {journal} {\bibinfo  {journal} {Neural
  Comput.}\ }\textbf {\bibinfo {volume} {15}},\ \bibinfo {pages} {509}
  (\bibinfo {year} {2003})}\BibitemShut {NoStop}%
\bibitem [{\citenamefont {Izhikevich}(2003)}]{Izhikevich_2003}%
  \BibitemOpen
  \bibfield  {author} {\bibinfo {author} {\bibfnamefont {E.~M.}\ \bibnamefont
  {Izhikevich}},\ }\href@noop {} {\bibfield  {journal} {\bibinfo  {journal}
  {IEEE Trans. Neural Netw.}\ }\textbf {\bibinfo {volume} {14}},\ \bibinfo
  {pages} {1569} (\bibinfo {year} {2003})}\BibitemShut {NoStop}%
\bibitem [{\citenamefont {Yuste}(2015)}]{Yuste_2015}%
  \BibitemOpen
  \bibfield  {author} {\bibinfo {author} {\bibfnamefont {R.}~\bibnamefont
  {Yuste}},\ }\href@noop {} {\bibfield  {journal} {\bibinfo  {journal} {Nat.
  Rev. Neurosci.}\ }\textbf {\bibinfo {volume} {16}},\ \bibinfo {pages} {487}
  (\bibinfo {year} {2015})}\BibitemShut {NoStop}%
\bibitem [{\citenamefont {Dorogovtsev}\ \emph {et~al.}(2002)\citenamefont
  {Dorogovtsev}, \citenamefont {Goltsev},\ and\ \citenamefont
  {Mendes}}]{Dorogovtsev_2002b}%
  \BibitemOpen
  \bibfield  {author} {\bibinfo {author} {\bibfnamefont {S.~N.}\ \bibnamefont
  {Dorogovtsev}}, \bibinfo {author} {\bibfnamefont {A.~V.}\ \bibnamefont
  {Goltsev}}, \ and\ \bibinfo {author} {\bibfnamefont {J.~F.~F.}\ \bibnamefont
  {Mendes}},\ }\href@noop {} {\bibfield  {journal} {\bibinfo  {journal} {Phys.
  Rev. E}\ }\textbf {\bibinfo {volume} {66}},\ \bibinfo {pages} {016104}
  (\bibinfo {year} {2002})}\BibitemShut {NoStop}%
\bibitem [{\citenamefont {{Dorogovtsev}}\ \emph {et~al.}(2008)\citenamefont
  {{Dorogovtsev}}, \citenamefont {{Goltsev}},\ and\ \citenamefont
  {{Mendes}}}]{Dorogovtsev_2008}%
  \BibitemOpen
  \bibfield  {author} {\bibinfo {author} {\bibfnamefont {S.~N.}\ \bibnamefont
  {{Dorogovtsev}}}, \bibinfo {author} {\bibfnamefont {A.~V.}\ \bibnamefont
  {{Goltsev}}}, \ and\ \bibinfo {author} {\bibfnamefont {J.~F.~F.}\
  \bibnamefont {{Mendes}}},\ }\href@noop {} {\bibfield  {journal} {\bibinfo
  {journal} {Rev. Mod. Phys.}\ }\textbf {\bibinfo {volume} {80}},\ \bibinfo
  {pages} {1275} (\bibinfo {year} {2008})}\BibitemShut {NoStop}%
\bibitem [{\citenamefont {Lopes}\ \emph {et~al.}(2016)\citenamefont {Lopes},
  \citenamefont {Lopes}, \citenamefont {Yoon}, \citenamefont {Mendes},\ and\
  \citenamefont {Goltsev}}]{Lopes_2016}%
  \BibitemOpen
  \bibfield  {author} {\bibinfo {author} {\bibfnamefont {M.~A.}\ \bibnamefont
  {Lopes}}, \bibinfo {author} {\bibfnamefont {E.~M.}\ \bibnamefont {Lopes}},
  \bibinfo {author} {\bibfnamefont {S.}~\bibnamefont {Yoon}}, \bibinfo {author}
  {\bibfnamefont {J.~F.~F.}\ \bibnamefont {Mendes}}, \ and\ \bibinfo {author}
  {\bibfnamefont {A.~V.}\ \bibnamefont {Goltsev}},\ }\href {\doibase
  10.1103/PhysRevE.94.012308} {\bibfield  {journal} {\bibinfo  {journal} {Phys.
  Rev. E}\ }\textbf {\bibinfo {volume} {94}},\ \bibinfo {pages} {012308}
  (\bibinfo {year} {2016})}\BibitemShut {NoStop}%
\bibitem [{\citenamefont {Giuraniuc}\ \emph {et~al.}(2006)\citenamefont
  {Giuraniuc}, \citenamefont {Hatchett}, \citenamefont {Indekeu}, \citenamefont
  {Leone}, \citenamefont {P\'erez~Castillo}, \citenamefont {Van~Schaeybroeck},\
  and\ \citenamefont {Vanderzande}}]{Giuraniuc_2006}%
  \BibitemOpen
  \bibfield  {author} {\bibinfo {author} {\bibfnamefont {C.~V.}\ \bibnamefont
  {Giuraniuc}}, \bibinfo {author} {\bibfnamefont {J.~P.~L.}\ \bibnamefont
  {Hatchett}}, \bibinfo {author} {\bibfnamefont {J.~O.}\ \bibnamefont
  {Indekeu}}, \bibinfo {author} {\bibfnamefont {M.}~\bibnamefont {Leone}},
  \bibinfo {author} {\bibfnamefont {I.}~\bibnamefont {P\'erez~Castillo}},
  \bibinfo {author} {\bibfnamefont {B.}~\bibnamefont {Van~Schaeybroeck}}, \
  and\ \bibinfo {author} {\bibfnamefont {C.}~\bibnamefont {Vanderzande}},\
  }\href {\doibase 10.1103/PhysRevE.74.036108} {\bibfield  {journal} {\bibinfo
  {journal} {Phys. Rev. E}\ }\textbf {\bibinfo {volume} {74}},\ \bibinfo
  {pages} {036108} (\bibinfo {year} {2006})}\BibitemShut {NoStop}%
\bibitem [{\citenamefont {Wang}\ \emph {et~al.}(1995)\citenamefont {Wang},
  \citenamefont {Golomb},\ and\ \citenamefont {Rinzel}}]{Wang_1995}%
  \BibitemOpen
  \bibfield  {author} {\bibinfo {author} {\bibfnamefont {X.}~\bibnamefont
  {Wang}}, \bibinfo {author} {\bibfnamefont {D.}~\bibnamefont {Golomb}}, \ and\
  \bibinfo {author} {\bibfnamefont {J.}~\bibnamefont {Rinzel}},\ }\href@noop {}
  {\bibfield  {journal} {\bibinfo  {journal} {Proc. Natl. Acad. Sci. USA}\
  }\textbf {\bibinfo {volume} {92}},\ \bibinfo {pages} {5577} (\bibinfo {year}
  {1995})}\BibitemShut {NoStop}%
\bibitem [{\citenamefont {Vreeswijk}\ and\ \citenamefont
  {Sompolinsky}(1998)}]{Vreeswijk_1998}%
  \BibitemOpen
  \bibfield  {author} {\bibinfo {author} {\bibfnamefont {C.~v.}\ \bibnamefont
  {Vreeswijk}}\ and\ \bibinfo {author} {\bibfnamefont {H.}~\bibnamefont
  {Sompolinsky}},\ }\href@noop {} {\bibfield  {journal} {\bibinfo  {journal}
  {Neural Comput.}\ }\textbf {\bibinfo {volume} {10}},\ \bibinfo {pages} {1321}
  (\bibinfo {year} {1998})}\BibitemShut {NoStop}%
\bibitem [{\citenamefont {Kadmon}\ and\ \citenamefont
  {Sompolinsky}(2015)}]{Kadmon_2015}%
  \BibitemOpen
  \bibfield  {author} {\bibinfo {author} {\bibfnamefont {J.}~\bibnamefont
  {Kadmon}}\ and\ \bibinfo {author} {\bibfnamefont {H.}~\bibnamefont
  {Sompolinsky}},\ }\href@noop {} {\bibfield  {journal} {\bibinfo  {journal}
  {Phys. Rev. X}\ }\textbf {\bibinfo {volume} {5}},\ \bibinfo {pages} {041030}
  (\bibinfo {year} {2015})}\BibitemShut {NoStop}%
\bibitem [{\citenamefont {Daqing}\ \emph {et~al.}(2011)\citenamefont {Daqing},
  \citenamefont {Kosmidis}, \citenamefont {Bunde},\ and\ \citenamefont
  {Havlin}}]{Daqing_2011}%
  \BibitemOpen
  \bibfield  {author} {\bibinfo {author} {\bibfnamefont {L.}~\bibnamefont
  {Daqing}}, \bibinfo {author} {\bibfnamefont {K.}~\bibnamefont {Kosmidis}},
  \bibinfo {author} {\bibfnamefont {A.}~\bibnamefont {Bunde}}, \ and\ \bibinfo
  {author} {\bibfnamefont {S.}~\bibnamefont {Havlin}},\ }\href@noop {}
  {\bibfield  {journal} {\bibinfo  {journal} {Nat. Phys.}\ }\textbf {\bibinfo
  {volume} {7}},\ \bibinfo {pages} {481} (\bibinfo {year} {2011})}\BibitemShut
  {NoStop}%
\bibitem [{\citenamefont {Watts}\ and\ \citenamefont
  {Strogatz}(1998)}]{Watts_1998}%
  \BibitemOpen
  \bibfield  {author} {\bibinfo {author} {\bibfnamefont {D.~J.}\ \bibnamefont
  {Watts}}\ and\ \bibinfo {author} {\bibfnamefont {S.~H.}\ \bibnamefont
  {Strogatz}},\ }\href@noop {} {\bibfield  {journal} {\bibinfo  {journal}
  {Nature}\ }\textbf {\bibinfo {volume} {393}},\ \bibinfo {pages} {440}
  (\bibinfo {year} {1998})}\BibitemShut {NoStop}%
\bibitem [{\citenamefont {Barahona}\ and\ \citenamefont
  {Pecora}(2002)}]{Barahona_2002}%
  \BibitemOpen
  \bibfield  {author} {\bibinfo {author} {\bibfnamefont {M.}~\bibnamefont
  {Barahona}}\ and\ \bibinfo {author} {\bibfnamefont {L.~M.}\ \bibnamefont
  {Pecora}},\ }\href@noop {} {\bibfield  {journal} {\bibinfo  {journal} {Phys.
  Rev. Lett.}\ }\textbf {\bibinfo {volume} {89}},\ \bibinfo {pages} {054101}
  (\bibinfo {year} {2002})}\BibitemShut {NoStop}%
\bibitem [{\citenamefont {Goltsev}\ \emph {et~al.}(2010)\citenamefont
  {Goltsev}, \citenamefont {de~Abreu}, \citenamefont {Dorogovtsev},\ and\
  \citenamefont {Mendes}}]{Goltsev_2010}%
  \BibitemOpen
  \bibfield  {author} {\bibinfo {author} {\bibfnamefont {A.~V.}\ \bibnamefont
  {Goltsev}}, \bibinfo {author} {\bibfnamefont {F.~V.}\ \bibnamefont
  {de~Abreu}}, \bibinfo {author} {\bibfnamefont {S.~N.}\ \bibnamefont
  {Dorogovtsev}}, \ and\ \bibinfo {author} {\bibfnamefont {J.~F.~F.}\
  \bibnamefont {Mendes}},\ }\href {\doibase 10.1103/PhysRevE.81.061921}
  {\bibfield  {journal} {\bibinfo  {journal} {Phys. Rev. E}\ }\textbf {\bibinfo
  {volume} {81}},\ \bibinfo {pages} {061921} (\bibinfo {year}
  {2010})}\BibitemShut {NoStop}%
\bibitem [{\citenamefont {Lee}\ \emph {et~al.}(2014)\citenamefont {Lee},
  \citenamefont {Lopes}, \citenamefont {Mendes},\ and\ \citenamefont
  {Goltsev}}]{Lee_2014}%
  \BibitemOpen
  \bibfield  {author} {\bibinfo {author} {\bibfnamefont {K.-E.}\ \bibnamefont
  {Lee}}, \bibinfo {author} {\bibfnamefont {M.~A.}\ \bibnamefont {Lopes}},
  \bibinfo {author} {\bibfnamefont {J.~F.~F.}\ \bibnamefont {Mendes}}, \ and\
  \bibinfo {author} {\bibfnamefont {A.~V.}\ \bibnamefont {Goltsev}},\
  }\href@noop {} {\bibfield  {journal} {\bibinfo  {journal} {Phys. Rev. E}\
  }\textbf {\bibinfo {volume} {89}},\ \bibinfo {pages} {012701} (\bibinfo
  {year} {2014})}\BibitemShut {NoStop}%
\bibitem [{\citenamefont {Holstein}\ \emph {et~al.}(2013)\citenamefont
  {Holstein}, \citenamefont {Goltsev},\ and\ \citenamefont
  {Mendes}}]{Holstein_2013}%
  \BibitemOpen
  \bibfield  {author} {\bibinfo {author} {\bibfnamefont {D.}~\bibnamefont
  {Holstein}}, \bibinfo {author} {\bibfnamefont {A.~V.}\ \bibnamefont
  {Goltsev}}, \ and\ \bibinfo {author} {\bibfnamefont {J.~F.~F.}\ \bibnamefont
  {Mendes}},\ }\href@noop {} {\bibfield  {journal} {\bibinfo  {journal} {Phys.
  Rev. E}\ }\textbf {\bibinfo {volume} {87}},\ \bibinfo {pages} {032717}
  (\bibinfo {year} {2013})}\BibitemShut {NoStop}%
\bibitem [{\citenamefont {Lopes}\ \emph {et~al.}(2014)\citenamefont {Lopes},
  \citenamefont {Lee}, \citenamefont {Goltsev},\ and\ \citenamefont
  {Mendes}}]{Lopes_2014}%
  \BibitemOpen
  \bibfield  {author} {\bibinfo {author} {\bibfnamefont {M.~A.}\ \bibnamefont
  {Lopes}}, \bibinfo {author} {\bibfnamefont {K.-E.}\ \bibnamefont {Lee}},
  \bibinfo {author} {\bibfnamefont {A.~V.}\ \bibnamefont {Goltsev}}, \ and\
  \bibinfo {author} {\bibfnamefont {J.~F.~F.}\ \bibnamefont {Mendes}},\
  }\href@noop {} {\bibfield  {journal} {\bibinfo  {journal} {Phys. Rev. E}\
  }\textbf {\bibinfo {volume} {90}},\ \bibinfo {pages} {052709} (\bibinfo
  {year} {2014})}\BibitemShut {NoStop}%
\bibitem [{\citenamefont {Lopes}\ \emph {et~al.}(2017)\citenamefont {Lopes},
  \citenamefont {Lee},\ and\ \citenamefont {Goltsev}}]{Lopes_2017}%
  \BibitemOpen
  \bibfield  {author} {\bibinfo {author} {\bibfnamefont {M.~A.}\ \bibnamefont
  {Lopes}}, \bibinfo {author} {\bibfnamefont {K.-E.}\ \bibnamefont {Lee}}, \
  and\ \bibinfo {author} {\bibfnamefont {A.~V.}\ \bibnamefont {Goltsev}},\
  }\href@noop {} {\bibfield  {journal} {\bibinfo  {journal} {Phys. Rev. E}\
  }\textbf {\bibinfo {volume} {96}},\ \bibinfo {pages} {062412} (\bibinfo
  {year} {2017})}\BibitemShut {NoStop}%
\bibitem [{\citenamefont {Faisal}\ \emph {et~al.}(2008)\citenamefont {Faisal},
  \citenamefont {Selen},\ and\ \citenamefont {Wolpert}}]{Faisal_2008}%
  \BibitemOpen
  \bibfield  {author} {\bibinfo {author} {\bibfnamefont {A.}~\bibnamefont
  {Faisal}}, \bibinfo {author} {\bibfnamefont {L.}~\bibnamefont {Selen}}, \
  and\ \bibinfo {author} {\bibfnamefont {D.~M.}\ \bibnamefont {Wolpert}},\
  }\href@noop {} {\bibfield  {journal} {\bibinfo  {journal} {Nat. Rev.
  Neurosci.}\ }\textbf {\bibinfo {volume} {9}},\ \bibinfo {pages} {292}
  (\bibinfo {year} {2008})}\BibitemShut {NoStop}%
\bibitem [{\citenamefont {Abramowitz}\ and\ \citenamefont
  {Stegun}(1970)}]{Abramowitz_1970}%
  \BibitemOpen
  \bibfield  {author} {\bibinfo {author} {\bibfnamefont {M.}~\bibnamefont
  {Abramowitz}}\ and\ \bibinfo {author} {\bibfnamefont {I.~A.}\ \bibnamefont
  {Stegun}},\ }\href@noop {} {\emph {\bibinfo {title} {Handbook of mathematical
  functions: with formulas, graphs, and mathematical tables}}}\ (\bibinfo
  {publisher} {Courier Dover Publications},\ \bibinfo {address} {Washington,
  D.C.},\ \bibinfo {year} {1970})\BibitemShut {NoStop}%
\bibitem [{\citenamefont {Maslov}\ and\ \citenamefont
  {Sneppen}(2002)}]{Maslov_2002}%
  \BibitemOpen
  \bibfield  {author} {\bibinfo {author} {\bibfnamefont {S.}~\bibnamefont
  {Maslov}}\ and\ \bibinfo {author} {\bibfnamefont {K.}~\bibnamefont
  {Sneppen}},\ }\href@noop {} {\bibfield  {journal} {\bibinfo  {journal}
  {Science}\ }\textbf {\bibinfo {volume} {296}},\ \bibinfo {pages} {910}
  (\bibinfo {year} {2002})}\BibitemShut {NoStop}%
\bibitem [{\citenamefont {Strogatz}(1994)}]{Strogatz_1994}%
  \BibitemOpen
  \bibfield  {author} {\bibinfo {author} {\bibfnamefont {S.~H.}\ \bibnamefont
  {Strogatz}},\ }\href@noop {} {\emph {\bibinfo {title} {{Nonlinear Dynamics
  And Chaos: With Applications To Physics, Biology, Chemistry, And
  Engineering}}}}\ (\bibinfo  {publisher} {{Perseus Books Group}},\ \bibinfo
  {address} {New York},\ \bibinfo {year} {1994})\BibitemShut {NoStop}%
\bibitem [{\citenamefont {Orlandi}\ \emph {et~al.}(2013)\citenamefont
  {Orlandi}, \citenamefont {Soriano}, \citenamefont {Alvarez-Lacalle},
  \citenamefont {Teller},\ and\ \citenamefont {Casademunt}}]{Orlandi_2013}%
  \BibitemOpen
  \bibfield  {author} {\bibinfo {author} {\bibfnamefont {J.~G.}\ \bibnamefont
  {Orlandi}}, \bibinfo {author} {\bibfnamefont {J.}~\bibnamefont {Soriano}},
  \bibinfo {author} {\bibfnamefont {E.}~\bibnamefont {Alvarez-Lacalle}},
  \bibinfo {author} {\bibfnamefont {S.}~\bibnamefont {Teller}}, \ and\ \bibinfo
  {author} {\bibfnamefont {J.}~\bibnamefont {Casademunt}},\ }\href@noop {}
  {\bibfield  {journal} {\bibinfo  {journal} {Nat. Phys.}\ }\textbf {\bibinfo
  {volume} {9}},\ \bibinfo {pages} {582} (\bibinfo {year} {2013})}\BibitemShut
  {NoStop}%
\bibitem [{\citenamefont {Stanley}(1987)}]{Stanley_1987}%
  \BibitemOpen
  \bibfield  {author} {\bibinfo {author} {\bibfnamefont {H.~E.}\ \bibnamefont
  {Stanley}},\ }\href@noop {} {\emph {\bibinfo {title} {Introduction to Phase
  Transitions and Critical Phenomena}}}\ (\bibinfo  {publisher} {Oxford
  University Press},\ \bibinfo {address} {London},\ \bibinfo {year}
  {1987})\BibitemShut {NoStop}%
\bibitem [{\citenamefont {{Dorogovtsev}}\ and\ \citenamefont
  {{Mendes}}(2002)}]{Dorogovtsev_2002}%
  \BibitemOpen
  \bibfield  {author} {\bibinfo {author} {\bibfnamefont {S.~N.}\ \bibnamefont
  {{Dorogovtsev}}}\ and\ \bibinfo {author} {\bibfnamefont {J.~F.~F.}\
  \bibnamefont {{Mendes}}},\ }\href {\doibase 10.1080/00018730110112519}
  {\bibfield  {journal} {\bibinfo  {journal} {Adv. Phys.}\ }\textbf {\bibinfo
  {volume} {51}},\ \bibinfo {pages} {1079} (\bibinfo {year}
  {2002})}\BibitemShut {NoStop}%
\bibitem [{\citenamefont {Sporns}\ \emph {et~al.}(2004)\citenamefont {Sporns},
  \citenamefont {Chialvo}, \citenamefont {Kaiser},\ and\ \citenamefont
  {Hilgetag}}]{Sporns_2004}%
  \BibitemOpen
  \bibfield  {author} {\bibinfo {author} {\bibfnamefont {O.}~\bibnamefont
  {Sporns}}, \bibinfo {author} {\bibfnamefont {D.~R.}\ \bibnamefont {Chialvo}},
  \bibinfo {author} {\bibfnamefont {M.}~\bibnamefont {Kaiser}}, \ and\ \bibinfo
  {author} {\bibfnamefont {C.~C.}\ \bibnamefont {Hilgetag}},\ }\href@noop {}
  {\bibfield  {journal} {\bibinfo  {journal} {Trends Cogn. Sci.}\ }\textbf
  {\bibinfo {volume} {8}},\ \bibinfo {pages} {418} (\bibinfo {year}
  {2004})}\BibitemShut {NoStop}%
\bibitem [{\citenamefont {Gal}\ \emph {et~al.}(2017)\citenamefont {Gal},
  \citenamefont {London}, \citenamefont {Globerson}, \citenamefont {Ramaswamy},
  \citenamefont {Reimann}, \citenamefont {Muller}, \citenamefont {Markram},\
  and\ \citenamefont {Segev}}]{Gal_2017}%
  \BibitemOpen
  \bibfield  {author} {\bibinfo {author} {\bibfnamefont {E.}~\bibnamefont
  {Gal}}, \bibinfo {author} {\bibfnamefont {M.}~\bibnamefont {London}},
  \bibinfo {author} {\bibfnamefont {A.}~\bibnamefont {Globerson}}, \bibinfo
  {author} {\bibfnamefont {S.}~\bibnamefont {Ramaswamy}}, \bibinfo {author}
  {\bibfnamefont {M.~W.}\ \bibnamefont {Reimann}}, \bibinfo {author}
  {\bibfnamefont {E.}~\bibnamefont {Muller}}, \bibinfo {author} {\bibfnamefont
  {H.}~\bibnamefont {Markram}}, \ and\ \bibinfo {author} {\bibfnamefont
  {I.}~\bibnamefont {Segev}},\ }\href@noop {} {\bibfield  {journal} {\bibinfo
  {journal} {Nat. Neurosci.}\ }\textbf {\bibinfo {volume} {20}},\ \bibinfo
  {pages} {1004} (\bibinfo {year} {2017})}\BibitemShut {NoStop}%
\bibitem [{\citenamefont {Kandel}\ \emph {et~al.}(2000)\citenamefont {Kandel},
  \citenamefont {Schwartz},\ and\ \citenamefont {Jessell}}]{Kandel_2000}%
  \BibitemOpen
  \bibfield  {author} {\bibinfo {author} {\bibfnamefont {E.~R.}\ \bibnamefont
  {Kandel}}, \bibinfo {author} {\bibfnamefont {J.~H.}\ \bibnamefont
  {Schwartz}}, \ and\ \bibinfo {author} {\bibfnamefont {T.~M.}\ \bibnamefont
  {Jessell}},\ }\href@noop {} {\emph {\bibinfo {title} {{Principles of neural
  science}}}}\ (\bibinfo  {publisher} {McGraw-Hill},\ \bibinfo {address} {New
  York},\ \bibinfo {year} {2000})\BibitemShut {NoStop}%
\bibitem [{\citenamefont {Mountcastle}(2003)}]{Mountcastle_2003}%
  \BibitemOpen
  \bibfield  {author} {\bibinfo {author} {\bibfnamefont {V.~B.}\ \bibnamefont
  {Mountcastle}},\ }\href {\doibase 10.1093/cercor/13.1.2} {\bibfield
  {journal} {\bibinfo  {journal} {Cereb. Cortex}\ }\textbf {\bibinfo {volume}
  {13}},\ \bibinfo {pages} {2} (\bibinfo {year} {2003})}\BibitemShut {NoStop}%
\bibitem [{\citenamefont {Okujeni}\ \emph {et~al.}(2017)\citenamefont
  {Okujeni}, \citenamefont {Kandler},\ and\ \citenamefont
  {Egert}}]{Okujeni_2017}%
  \BibitemOpen
  \bibfield  {author} {\bibinfo {author} {\bibfnamefont {S.}~\bibnamefont
  {Okujeni}}, \bibinfo {author} {\bibfnamefont {S.}~\bibnamefont {Kandler}}, \
  and\ \bibinfo {author} {\bibfnamefont {U.}~\bibnamefont {Egert}},\
  }\href@noop {} {\bibfield  {journal} {\bibinfo  {journal} {J Neurosci.}\
  }\textbf {\bibinfo {volume} {37}},\ \bibinfo {pages} {3972} (\bibinfo {year}
  {2017})}\BibitemShut {NoStop}%
\end{thebibliography}%
\end{document}